\documentclass[12pt,a4paper]{article}
\title{
SU(2) Skyrme Vortices}
\author{
Yu.~P.~Rybakov, A.~M.~Tarabay, and I.~G.~Chugunov
\\
Department of Theoretical Physics, \\
Peoples' Friendship University of Russia, \\
6, Miklukho-Maklay Str., 117198 Moscow, Russia.\\
E-mail: yrybakov@mx.pfu.edu.ru
}

\date{}

\begin{document}
\maketitle

\begin{abstract}
A regular method for constructing vortex-like solutions with
cylindrical symmetry to the equations of the SU(2) Skyrme chiral model is
proposed. A numerical estimate for the length density of mass is given.
\end{abstract}

The Skyrme model [1] proved its efficiency in modelling the structure of
baryons [2] and nuclei [3] since the appearance of Witten's analysis
of quark    confinement   problem[4,5].
The model considers pions as Goldstone bosons and uses the Lagrangian
density
\begin{equation}
{\cal L}=-\frac{1}{4\lambda^{2}}{\rm
Sp}\,(l_{\mu}l^{\mu})+\frac{\varepsilon^2}{16}{\rm
Sp}([l_\mu,l_\nu][l^\mu,l^\nu]),
\end{equation}
constructed  from  the  chiral  current $l_\mu=U^+\partial_\mu
U,\,\,U\in SU(2).$ The energy in the model is estimated from below through
the topological charge
$$
Q=-\frac{1}{24\pi^2}\varepsilon^{ijk}\int_{}^{}d^3      x\,{\rm
Sp}(l_i\,l_j\,l_k)
$$
which takes integer values and can be interpreted as baryon number. In
particular, the nucleon emerges as an absolutely stable state with the
minimal energy in the first homotopic class $(Q=1)$ [6]. Unfortunately
the corresponding hedgehog configuration cannot be described analytically
due to complexity of the nonlinear equations for the chiral field.
The situation is aggravated for the higher homotopic classes in view of
nonseparability of radial and angle variables. To overcome these difficulties
we propose to approximate the configurations with higher charges by closed
vortices. As a first step in this direction we consider in the present paper
the simplest static vortex configuration given by the matrix
\begin{equation}
U=\exp\,(i\tau\Theta(\rho)),\quad
\tau=\left[\matrix{0& e^{-i\varphi}\cr e^{i\varphi}&0\cr}\right],
\end{equation}
with  $\rho,\,\varphi$ being cylindrical coordinates.
The  configuration  (2)  appears  to  be equivariant under the
group $G=T(z)\otimes {\rm diag}[{SO(2)}_I\otimes {SO(2)}_S]$
including the translation along the vortex and combined isotopic-space
rotations around its axis. Usually the massive term 
proportional to $2 - U - U^{+}$ is added
to the expression (1). Though its role being essential for the existence of 
regular solutions, we omit it 
in the present text, with the aim to illustrate 
the method and to simplify the formulae.

Substituting (2) in (1)  amounts to the radial 
Lagrangian density for the chiral
angle  $\Theta(\rho)$:
$$
{\cal
L}=-\frac{1}{2\lambda^2}\left({\Theta'}^2+\frac{\sin^2\Theta}{\rho
^2}\right)-\varepsilon^2{\Theta'}^2\frac{\sin^2\Theta}{\rho^2}.
$$

After  the  change of variable $\rho=\lambda\sqrt 2 e^{t},\,\,
-\infty\le t \le +\infty,\,\,$    we obtain the mechanical problem given by
the action functional
\begin{equation}
I[\Theta]=\int\limits_{-\infty}^{+\infty}dt\left[
{\dot                                \Theta}^2(
1+\varepsilon^2 e^{-2t}\sin^2\Theta)+\sin^2\Theta
\right],
\end{equation}
where the dimensionless parameter   $\varepsilon^2$  is reserved for the technical
purposes.

From (3) one derives the canonical momentum
\begin{equation}
p=2{\dot \Theta}{(
1+\varepsilon^2
e^{-2t}\sin^2\Theta)}=\partial_\Theta S
\end{equation}
and the Hamilton-Jacobi equation
\begin{equation}
\partial_t S+\frac{1}{4}(\partial_\Theta S)^2 {(1+\varepsilon^2
e^{-2t}\sin^2\Theta)}^{-1}-\sin^2\Theta=0.
\end{equation}

Now we search for the solution to the Eq.(5) as a formal series
\begin{equation}
S(t,\,\Theta)=
\sum_{n=0}^{\infty}\varepsilon^{2n}S_n(t,\,\Theta).
\end{equation}

Inserting (6) into (5) we get the recurrence relation
\begin{equation}
\frac{1}{4}\sum_{l=0}^{n}\partial_\Theta
S_l\,\partial_\Theta     S_{n-l}+\partial_t     S_n+\partial_t
S_{n-1}e^{-2t}\sin^2\Theta= \\      \delta_{n0}\sin^2\Theta+
\delta_{n1}e^{-2t}\sin^4\Theta
\end{equation}
with the evident private solution for $n=0:$
\begin{equation}
S_0=W_0(\Theta)=2\cos\Theta.
\end{equation}

In order to satisfy (7) we put for $n>0$
\begin{equation}
S_n(t,\,\Theta)=e^{-2nt}W_n(\Theta).
\end{equation}

From (7), (8) and (9) we derive for $n=1$ the equation
$$
\sin\Theta\,W_1'+2W_1+\sin^4\Theta=0
$$
with the two different solutions (two branches)
corresponding   to   the   ranges  $\pi/2\le\Theta\le\pi$  and
$0\le\Theta\le\pi/2$ respectively:
\begin{equation}
W_1^+(\Theta)=\frac{1}{3}(1+\cos\Theta)^2\left(
\frac{4}{1-\cos\Theta}+3-\cos\Theta
\right),
\end{equation}
\begin{equation}
W_1^-(\Theta)=-\frac{1}{3}(1-\cos\Theta)\sin^2\Theta.
\end{equation}

For $n>1$ we deduce from (7), (8) and (9) the equation
$$
\begin{array}{l}
\sin\Theta  W_n'+2nW_n=
\frac14\sum_{l=1}^{n-1}W_l'W_{n-l}'-2
(n-1)\sin^2\Theta W_{n-1}
\end{array}
$$
with the solution
\begin{equation}
\begin{array}{l}
W_n^{\pm}(\Theta)=
{\rm tg}^{-2n}(\Theta/2)\int\limits_{a_{\pm}}^{\Theta
}\frac{d\Theta}{\sin\Theta}{\rm tg}^{2n}(\Theta/2)\times{} \qquad {}
\\[10pt]
\qquad
\left[
\frac14\sum_{l=1}^{n-1}W_l'W_{n-l}'-2
(n-1)\sin^2\Theta W_{n-1}
\right],
\end{array}
\end{equation}
where $a_+=\pi,\,\,a_-=0.$ Formulae (10), (11) and (12) determine the recurrent procedure
for constructing the solution to the Hamilton-Jacobi equation (5).
Putting this solution to the r.h.s. of the equation (4), one can find the
canonical momentum $p$ for the two branches of solution:
\begin{equation}
p=2{\dot \Theta}{(
1+\varepsilon^2
e^{-2t}\sin^2\Theta)}=\sum_{n=0}^{\infty}\varepsilon^{2n}e^{-2
nt}W_n'(\Theta).
\end{equation}

For  matching  these  branches at the point $\Theta=\pi/2,\,\,
t=t_0$ one deduces from (13)
the algebraic equation
\begin{equation}
\sum_{n=1}^{\infty}\varepsilon^{2n}e^{-2nt_0}[{W_n^{+}}'(\pi/2)-
{W_n^{-}}'(\pi/2)]=0.
\end{equation}

In particular, within the scope of $n=2$ approximation
one gets from (14) the effective development parameter
$$
\xi=\frac13\varepsilon^2  e^{-2t_0}=2(16\ln 2-37/15)^{-1}\approx
0,232
$$
that  permits  to calculate the mass of the vortex (its length
density):

$$
\begin{array}{l}
M=\frac\pi{\lambda^2}I
=\frac\pi{\lambda^2}[
S^-(+\infty,\, 0)-S^-(t_0,\,\pi/2)+
S^+(t_0,\,\pi/2)-S^+(-\infty,\pi)]=
 \\[10pt]
\frac{4\pi}{\lambda^2}[1+2\xi+\xi^2(41/15-8\ln
2)+O(\xi^3)]\approx\frac{4\pi}{\lambda^2}1,31.\end{array}
$$

In conclusion we notice that the function  $\Theta(t),$ defining the radial distribution
of matter inside the vortex, can be found from Eq.(13) which is represented
in the integral form:
$$
\begin{array}{l}
{\rm tg}\frac\Theta 2=
e^{t_0 - t }\exp\left\{\int\limits_{t_0}^t         dt(1+\varepsilon^2
e^{-2t}\sin^2\Theta)^{-1}\times \right.\\{}
\left.
\left(
\varepsilon^2     e^{-2t}\sin^2\Theta+\frac12\sum\limits_{n=1}^\infty
\varepsilon^{2n}e^{-2nt}\frac{W_n'}{\sin\Theta}\right)\right\}.
\end{array}
$$

\end{document}